\pdfoutput=1
\documentclass[aps,prb,twocolumn,showpacs,superscriptaddress,amssymb,preprintnumbers]{revtex4}
\usepackage{graphicx}
\usepackage{bm}                   
\usepackage{amsmath}
\usepackage{dcolumn}        
\usepackage[caption=false]{subfig}
\usepackage[dvipsnames]{xcolor}				
\usepackage[linktocpage, colorlinks, linkcolor = blue, citecolor = blue]{hyperref}
\begin{document}
	
	\title{Low temperature features in the heat capacity of unary metals and intermetallics for the example of bulk aluminum and Al$_3$Sc}
	
	\author{Ankit Gupta}
	\affiliation{Max-Planck-Institut f\"ur Eisenforschung GmbH, D-40237 D\"usseldorf, Germany}
	\author{Beng\"u Tas Kavakbasi}
	\affiliation{Institute of Materials Physics, University of M\"unster, D-48149 M\"unster, Germany}
	\author{Biswanath Dutta}
	\affiliation{Max-Planck-Institut f\"ur Eisenforschung GmbH, D-40237 D\"usseldorf, Germany}
	\author{Blazej Grabowski}
	\affiliation{Max-Planck-Institut f\"ur Eisenforschung GmbH, D-40237 D\"usseldorf, Germany}
	\author{Martin Peterlechner}
	\affiliation{Institute of Materials Physics, University of M\"unster, D-48149 M\"unster, Germany}
	\author{Tilmann Hickel}
	\affiliation{Max-Planck-Institut f\"ur Eisenforschung GmbH, D-40237 D\"usseldorf, Germany}
	\author{Sergiy V. Divinski}
	\affiliation{Institute of Materials Physics, University of M\"unster, D-48149 M\"unster, Germany}
	\author{Gerhard Wilde}
	\affiliation{Institute of Materials Physics, University of M\"unster, D-48149 M\"unster, Germany}
	\author{J\"org Neugebauer}
	\affiliation{Max-Planck-Institut f\"ur Eisenforschung GmbH, D-40237 D\"usseldorf, Germany}
	\date{\today}
	
	\begin{abstract}
		We explore the competition and coupling of vibrational and electronic contributions to the heat capacity of Al and Al$_3$Sc at temperatures below 50 K combining experimental calorimetry with highly converged finite temperature density functional theory calculations. We find that semilocal exchange correlation functionals accurately describe the rich feature set observed for these temperatures, including electron-phonon coupling. Using different representations of the heat capacity, we are therefore able to identify and explain deviations from the Debye behaviour in the low-temperature limit and in the temperature regime 30 -- 50 K as well as the reduction of these features due to the addition of Sc. 
		
	\end{abstract}
	
	\maketitle
	
	\section{Introduction}
	
	The heat capacity is one of the key physical quantities in materials research, since it can be directly measured and has a significant technological and scientific impact. It is crucial to determine the onset of phase transitions~\cite{PhaseTransition1,PhaseTransition2} and to construct thermodynamic data bases (e.g., using the CALPHAD technique\cite{Kaufman}), to evaluate magnetocaloric effects~\cite{Magnetocaloric1,Magnetocaloric2}, as well as for applications in microelectronics~\cite{Microelectronics1} and modern battery materials~\cite{Lithiumion1}. Computing the heat capacity and understanding its generic features requires an accurate determination of the temperature dependence of all relevant excitation mechanisms such as atomic vibrations, electronic excitations and the coupling between them.
	
	Recent advances in finite-temperature first-principles calculations have led to an accurate description of key thermodynamic quantities of a broad range of material systems~\cite{OzoAsta2001solvus,FritzIron2012,Previous12,Previous7,ZugangMao2011,quasiharm1,quasiharm2,quasiharm3,BiswanathPRL}. With the \textit{ab initio} computed Gibbs energy surface at hand, all measurable thermodynamic properties such as e.g. thermal expansion or temperature dependent elastic constants for a given compound can be accessed. Out of this set, the heat capacity $C_P$ is particularly sensitive to fine details in the excitation mechanisms, since it is a second derivative of the Gibbs free energy. Special care is required at low temperatures -- as demonstrated in the present work -- where only the consideration of $C_P/T^n$ with $n=0, \ldots, 3$ allows a full evaluation of thermodynamic contributions to the heat capacity in non-magnetic metals\cite{footnoteMagnetic} and alloys.  
	
	The Debye model is commonly used to describe the temperature dependence of the heat capacity of materials at $T\ll \Theta_D$,  ($\Theta_D$ : Debye temperature). According to this model the heat capacity should exhibit a $T^3$ dependence. Experimentally, however, in many metals deviations from the $T^3$ behaviour are observed at temperatures below 50 K. These deviations can be attributed to the realistic phonon dispersion as well as to electronic excitations close to the Fermi surface~\cite{KittelBook,IIScBook,AlumCP4K}.
	Recent finite-temperature density functional theory (DFT) calculations have indeed shown the importance of electronic contributions to the heat capacity for various metals~\cite{Previous1,Previous2,Previous3,Previous4,Previous5,Previous6,Previous7,Previous8,Previous9,Previous10,Previous11,Previous12,Previous13,Previous14}. In all these works, however, the focus was on the high temperature regime ($T>\Theta_D$), where the importance of the electronic contributions also intuitively seems to be highest. While temperature effects well below $\Theta_D$ are small on an absolute scale  and have thus little effect on materials properties,  analyzing  the asymptotic behavior of the heat capacity towards absolute zero promises detailed insights into the various excitation mechanisms. The careful analysis of the low-temperature regime, both with respect to the methodological tools as well to the physical phenomena,  is therefore in the focus of the present study.
	
	Expressing calorimetric data solely as a sum of a vibrational phonon contribution (cubic in $T$) and an electronic contribution (linear in $T$)~\cite{KittelBook,IIScBook} neglects couplings between these excitation mechanisms. The interplay between electronic and vibrational degrees of freedom yields adiabatic as well as non-adiabatic corrections. The impact of electronic temperature on the interatomic forces and thus on phonon energies is typically added to the adabiatic part. The term ``electron-phonon coupling'' is (in the context of perturbation theory) reserved for transition probabilities between electronic states due to ionic displacements \cite{BWC06}. It is described within the Eliashberg theory by a coupling strength, which rescales the electronic heat capacity by a mass enhancement parameter\cite{BWC06,BSP98,Gri76,HEC02,SS96}. It is directly related to the phonon linewidth arising from interactions with electrons\cite{BSP98}. 
	While electron-phonon coupling is an established concept in the field of superconductivity, it is rarely discussed in calorimetry. Only few studies used this contribution to explain observed deviations in heat capacity measurements of simple metals when using only the two adiabatic contributions. In contrast to the present study the focus in these works was on temperatures well above the Debye temperature \cite{LIN,WSB16}. 
	Though it is mathematically clear that the linear (i.e. electronic) term will dominate over the cubic (i.e. vibrational) term at sufficiently low temperatures, a detailed analysis how well DFT captures this competition and the coupling of electronic and vibrational degrees of freedom at very low temperatures is currently missing. Such a benchmark is in particular important for alloys, for which  low temperature experimental data are typically not readily available, making finite-temperature DFT calculations a promising tool for an accurate assessment of physical properties.
	
	In the present work, we investigate the aforementioned deviations from the Debye behaviour in the low temperature regime within a full \textit{ab initio} approach and perform experiments as a benchmark. We  take pure Al and the intermetallic compound Al$_3$Sc as prototypical systems. The Al$_3$Sc precipitate phase is of prime importance to achieve the high mechanical strength of Al rich Al-Sc alloys and has, over the last couple of decades, enabled the impressive development of these alloys~\cite{Nature1,Nature2,Parker1995,Drits1984,Elagin1992,Rostova2000,Norman1998,Milman2000,Nature3,Nature4,Harada2002,Hyland1992,Harada2003,Ahmad2003,AstaOzo2001properties,AstaFoiles1998,MarquisSeidman2001,MarquisAsta2006,DunandSeidman2008,OzoAsta2001solvus,RoysetRyum2005ScinAl,ZugangMao2011,ClouetMartin2005,ClouetSigliAstaFoiles1998,Saha2015}. Besides possessing a structural (fcc crystal structure) and dimensional  (lattice mismatch approx.~1.6\% at 0 K, decreasing with temperature) correspondence with the Al host matrix, these precipitates are coherently incorporated in the Al host matrix and homogeneously distributed resulting in a fine-grained microstructure, as well as high yield and tensile strength. Since both, Al and Al$_3$Sc, share a common underlying face-centered cubic (fcc) lattice, with Sc substituting the corners of the conventional cubic unit cell, we can directly connect the observed changes in the heat capacity features with the chemical impact of Sc. 
	
	\section{Methodology}
	
	To derive the thermodynamic properties of a given material system within DFT, the Helmholtz free energy is a common starting point. For a defect free non-magnetic material, the free energy includes the following  contributions,
	\begin{align}
	F(V,T) &= E_{0\text{K}}(V) + F^{\text{el}}(V,T) + F^{\text{qh}}(V,T) \nonumber\\
	&\quad + F^{\text{ph-ph}}(V,T)   + F^{\text{el-ph}}(V,T),  \label{eq:1} 
	\end{align}
	where the electronic (el), quasiharmonic (qh) and the anharmonic phonon-phonon coupling (ph-ph) constitute the adiabatic approximation~\cite{Previous13}, while the electron-phonon coupling (el-ph) contains also non-adiabatic contributions. 
	The target isobaric heat capacity $C_P$ is derived from the free energy using the relation,
	\begin{equation}
	C_P=-T\left(\frac{\partial^2 F(V,T)}{\partial T^2}\right)_{V,P}.
	\end{equation} 
	In the following the employed methods to calculated the different contributions to Eq. (\ref{eq:1}) are briefly discussed.
	
	\subsection{Electronic contribution}
	The electronic contribution $ F^{\text{el}}(V,T)$, which is particularly important for the present low temperature study, is obtained following Mermin's finite-temperature DFT formalism~\cite{Mermin,Previous13}. Its physical origin is the thermodynamic excitation of electrons close to the Fermi energy (within a range of approx.~$k_\text{B}T$), which yields in lowest order of the Sommerfeld expansion for the ideal Thomas Fermi gas a temperature dependence\cite{KittelBook,IIScBook}
	\begin{equation}
	C_P^{\text{el}}(T) = \gamma T  \label{eq:Sommerfeld}
	\end{equation}
	with  the Sommerfeld coefficient $\gamma$.
	
	The $\{V,T\}$ parametrization of $ F^{\text{el}}(V,T)$ is done as follows: Firstly, we calculate the $T$-dependent part of $ F^{\text{el}}$ by performing DFT calculations on an equidistant mesh of 11 volumes and 11 temperatures as
	\begin{equation}
	\label{eq:2}
	F^{\text{el}}(V,T ) = F^{\text{el}}_\text{tot} (V,T ) - E_{\text{0K}}(V),
	\end{equation} 
	where $F^{\text{el}}_\text{tot}$ is the total electronic free energy (including the 0 K binding energy $E_{\text{0K}}$) obtained from a fully self-consistent DFT calculation at a finite electronic temperature $T$ corresponding to a certain Fermi smearing. To obtain a dense temperature sampling for calculating the heat capacity, we use a physically motivated fit \cite{Previous13}, $F^{\text{el}}(T)= -\frac{1}{2}T S^{\text{el}}(T)$, where 
	\begin{equation}
	\label{eq:3}
	S^{\text{el}}(T)=-2k_B\int d\epsilon \:  N^{\text{el}}(T)\: [f \ln f + (1-f)\: \ln(1-f)],
	\end{equation}
	with the Fermi-Dirac distribution function $f=f(\epsilon,T)$ and with $N^{\text{el}}(T)$ representing an energy independent electronic density of states. The latter is used as a fitting quantity by expanding it up to a third-order polynomial in $T$ as $\sum_{i=0}^3 c_i T^i$ with fitting coefficients $c_i$. Next, for the parametrization of the volume dependence of $F^{\text{el}}$, a second-order polynomial is used to fit $F^{\text{el}}(V)$. The volume parametrization of $F^{\text{el}}$ is rather simple as we separated out the $T$ = 0 K binding energy $E_{\text{0K}}$, which carries a stronger volume dependence (fitted by a Murnaghan equation of state~\cite{Murn}). In this way a numerical error of less than 0.1 meV/atom is achieved at all temperatures and volumes. For further details, we refer to Ref.~\onlinecite{Previous13}.
	
	\subsection{Vibrational contributions} \label{sc:Vibrational}
	In Eq.~(\ref{eq:1}), the terms $F^{\text{qh}}$ and  $F^{\text{ph-ph}}$ together characterize the vibrational free energy. The quasiharmonic contribution, $F^{\text{qh}}(V,T)$, describes non-interacting,  volume dependent phonons and can straightforwardly be calculated~\cite{WallaceBook,Previous13}.  
	The above mentioned Debye approximation for the harmonic lattice vibrations yields the expression\cite{KittelBook}
	\begin{equation}
	C_V^{\text{h}} = \frac{12 \pi^4 k_\text{B}}{5} \left( \frac{T}{\Theta_D} \right)^3
	\label{eq:Debye}
	\end{equation}
	for the heat capacity, which is cubic in $T$.
	
	To estimate the influence of electronic temperature on the phonon frequencies, we have calculated the variation of the highest optical phonon frequency for Al$_3$Sc (marked by an orange arrow at point R in Fig.~\ref{dos_and_phonon}c) with the Fermi smearing width $\sigma$ ranging from 0.01 eV to 0.15 eV (120 K -- 1740 K). The variation of this phonon frequency is on the order of $2.4\times 10^{-2}$ THz, which  corresponds to an energy change of  $\approx0.1$ meV. We have determined the impact of the resulting temperature dependence on the heat capacity and found it to be an order of magnitude smaller than the contribution of $F^{\text{qh}}$. 
	
	In addition we have calculated the modification of the heat capacity due to explicitly anharmonic vibrations caused by phonon-phonon interaction. In accordance with previous studies\cite{Previous8}, this effect also turns out to be negligible for the temperature regime investigated in this work ($< 400$ K). Having in mind that already the quasiharmonic contribution $F^{\text{qh}}$ is for the critical temperature regime ($< 5$ K) much smaller than the electronic contribution $F^{\text{el}}$, we do not include the modification of lattice vibrations due to electronic temperature or phonon-phonon interaction in the upcoming discussion.  
	
	\subsection{Electron-phonon coupling}\label{sssec:El-Ph-Coupling}
	The electron-phonon coupling has been shown~\cite{Gri76} to be crucial at very low temperatures (typically 0.1--4 K) for describing the deviation of the electronic heat capacity $C_P^{\text{el}}$ from the linear temperature dependence according to Eq.~(\ref{eq:Sommerfeld}) that is solely captured by the Sommerfeld coefficient  $\gamma$ . Considering the coupling, the electronic heat capacity is written as 
	\begin{equation}
	\label{eq:epcoupling}
	C_P^{\text{el}} + C_P^{\text{el-ph}} = \gamma (1+\lambda) T \quad , 
	\end{equation}
	where $\lambda$ is the dimensionless electron-phonon coupling parameter. This parameter is formally defined as the first reciprocal moment of the Eliashberg spectral function\cite{BSP98,Gri76,HEC02,SS96} $\alpha^2 g(\nu)$,
	\begin{equation}
	\label{eq:eliashberg}
	\lambda = 2 \int_{0}^{\infty}\frac{\alpha^2 g(\nu)}{\nu} d\nu \quad .
	\end{equation} 
	The Eliashberg spectral function represents the vibrational density of states (VDOS) $g(\nu)$ weighted by the effective electron-phonon coupling function averaged over the Fermi surface $\alpha^2$ as explained in Refs.~\onlinecite{BS54,SS96,BSP98}. 
	
	Physically, it quantifies the contribution of phonons with frequency $\nu$ to the scattering process of electrons at the Fermi level\cite{BSP98}. The energy changes due to this coupling can be obtained from diagrammatic perturbation theory. The resulting temperature dependence is mainly determined by a product of Fermi distribution and Bose-Einstein distribution functions. In the low temperature limit, however, only the $T$ independent zero point vibrations are relevant\cite{BS54,BWC06}, yielding an overall linear $T$ dependence of $C_P^{\text{el-ph}}$ similar to the Sommerfeld expansion.  Further studies\cite{LIN,WSB16} demonstrate that the coupling affects electronic and vibrational degrees of freedom simultaneously.
	
	\subsection{Computational Details} \label{sc:CompDetails}
	Total energy and force calculations are performed for a $4\times4\times4$ fcc supercell within the  projector-augmented wave (PAW) method\cite{PAW} as implemented in the Vienna \emph{Ab Initio} Simulation Package ({\sc vasp})~\cite{Kresse1993,Kresse1996}. The generalized gradient approximation (GGA) as parameterized by Perdew-Burke-Ernzerhof (PBE)\cite{PBE} is used for the exchange-correlation functional. In the case of Al, a comparison with the local density approximation (LDA) has been done to estimate the impact which the choice of the exchange-correlation functional  has on the results. As discussed in previous studies\cite{Previous7} these deviations can be used as an approximate DFT error bar.  
	
	The phonon calculations are performed employing the small displacement method\cite{KresseSmallDisplacement1995,Previous13} with a displacement value of 0.02 Bohr radius ($\approx$ 0.01\AA). The small value ensures that the forces on all the atoms within the given supercell vary linearly with the displacement. After performing convergence tests, a plane-wave energy cutoff \emph{E}$_{\rm cut}$ = 400 eV is used. The Monkhorst-Pack scheme~\cite{Monkhorst1976} for the sampling of the Brillouin zone (BZ) is chosen, with a reciprocal-space $k$-mesh of $6\times6\times6$ grid points. An energy of $10^{-7}$ eV is used as a convergence criterion for the self-consistent electronic loop. A Methfessel-Paxton scheme\cite{MP} with a width of 0.15 eV is used for the force calculations entering the dynamical matrix. 
	
	The frequencies needed to calculate $F^{\text{qh}}$ are generated by sampling the complete BZ using a q-mesh of $30\times30\times30$ grid points. A relative shift of $(0.25,0.25,0.25)$ with respect to the $\Gamma$ point is introduced. This is important for the calculation of the heat capacity in the limit $T \rightarrow$ 0 K, where only the modes in the vicinity of the $\Gamma$ point are excited, while the $\Gamma$ point itself needs to be excluded from the partition sum due to its diverging contribution. Even with the shift an artifical dependence on the choice of  $q$-mesh employed for sampling the BZ is expected, which makes the temperature region ($T\rightarrow0$) unreliable. We find that with a $4\times4\times4$ fcc supercell and an interpolation mesh of $30\times30\times30$ grid points with the $(0.25,0.25,0.25)$ shift, the Debye behavior is valid down to 2 K. By increasing the mesh size, for example, to $50\times50\times50$ one can extend the Debye behavior to below 1.8 K, but it would add to the computation time dramatically. Since deviations of the experimental $C_P/T^3$ data from linearity are already observed in the region between 2 K and 20 K (gray shaded regions in Figs.~\ref{sfig:cp_by_T3_al}), the lower limit of 2 K is sufficient for the purpose of the present study.
	
	For the calculations of the electronic free energy, a Fermi smearing with a width ranging from $8.6 \times 10^{-5}$ eV to 0.137 eV is used, corresponding to a temperature range of 1-1590 K (the upper limit corresponds to the melting temperature for Al$_3$Sc) with a mesh of 11 separate temperature values. The calculations are performed for a unit cell with a k-mesh of $20\times20\times20$ grid points  and a plane-wave energy cutoff of 300 eV.
	
	\begin{figure}[h]
		\includegraphics[scale=.45]{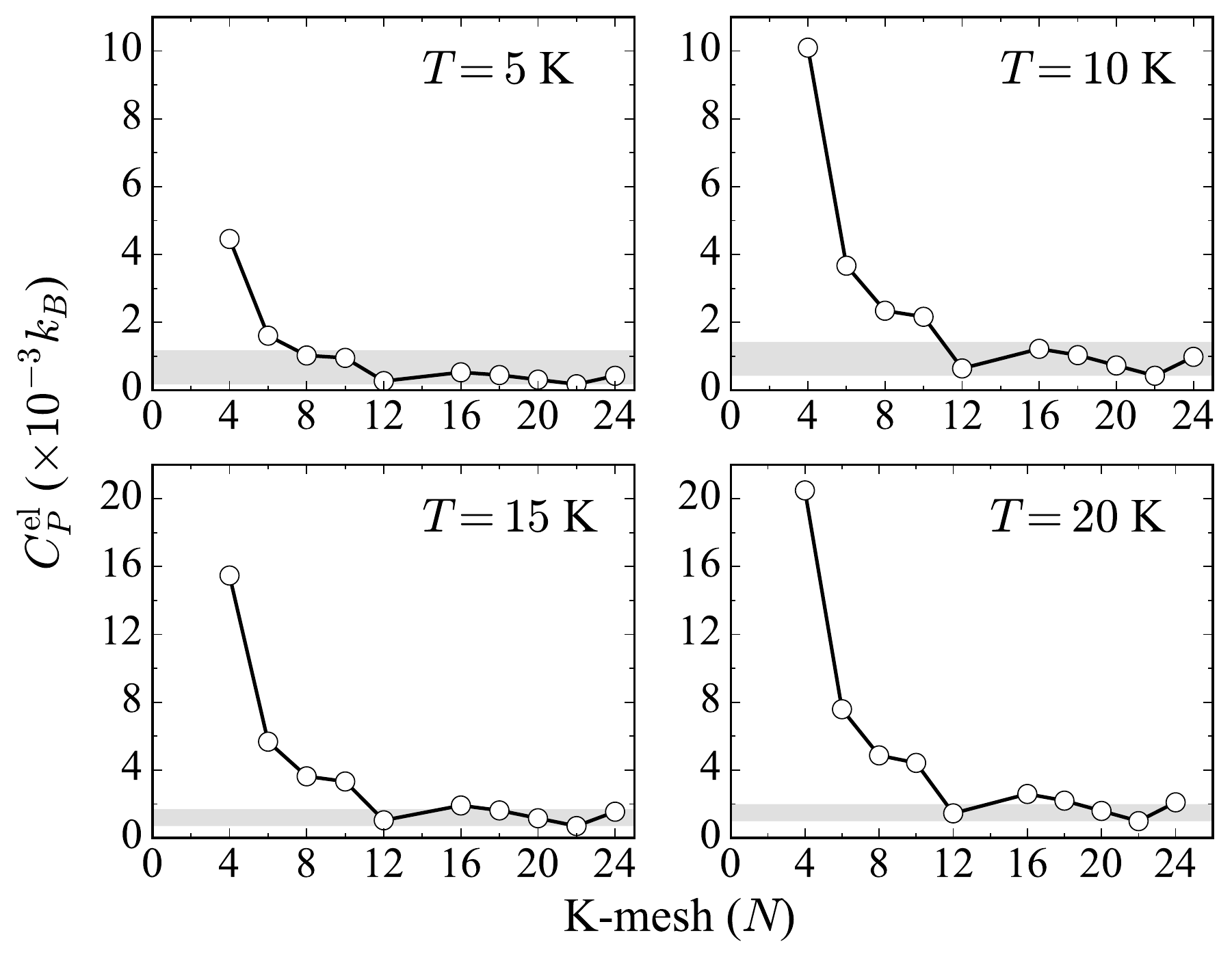}
		\caption{\label{fig1} Convergence of the $k$-mesh  for the electronic contribution to the heat capacity ($C_P^{\text{el}}$) in $k_B$ for $T$ = 5, 10, 15 and 20 K. The $N$ on the x-axis denotes the $k$-mesh size ($N^3$). The gray shaded area corresponds to a width of magnitude $1\times10^{-3} k_B$ (boundary values depend on temperature) within which the converged $C_P^{\text{el}}$ fluctuates.}
	\end{figure}
	
	Figure~\ref{fig1} shows the convergence of the electronic contribution to the heat capacity $C_P^{\text{el}}$ with respect to the $k$-mesh at the electronic temperature $T$ = 5, 10, 15 and 20 K. For $T= 15$ K, for example, a mesh of at least $12\times12\times12$ $k$-points is required to ensure that the error bar of $C_P^{\text{el}}$ is on the order of 0.001 $k_B$ (gray shaded regions in Fig.~\ref{fig1}). Since the $C_P^{\text{el}}$ values scale with $T$, one needs at higher temperatures larger $k$-meshes to match the same absolute error. For the present work, we chose a mesh of $20\times20\times20$ $k$-points. 
	
	The calculations to obtain the electron-phonon coupling parameter given by Eq. (\ref{eq:eliashberg}) are performed employing density functional perturbation theory (DFPT)\cite{ABINIT3} with PAW potentials (not identical to those in VASP) as implemented in the ABINIT code\cite{ABINIT1,ABINIT2}. For the exchange-correlation potential, we used the LDA as parametrized by Perdew and Wang\cite{LDAPW}, since DFPT is in ABINIT currently not implemented for GGA. The chosen values for the energy cutoff and the k-point grid are 20 Ha ($\approx$ 544 eV) and $28\times28\times28$, respectively. The calculations for Al$_3$Sc are performed with the equlilibrium lattice parameter of 7.62 Bohr (4.032 $\text{\AA}$). The LDA lattice constant obtained with VASP is 4.033 $\text{\AA}$, i.e. almost same as the value for the ABINIT pseudopotential. The convergence criterion for the self-consistent field cycle is 10$^{-14}$ Ha on the wave function squared residual.
	
	\subsection{Experimental Details}
	
	An Al$_3$Sc ingot was prepared by induction melting and casting into a copper mold from pure Al ($99.999$~wt.\%) and Sc ($99.99$~wt.\%) under vacuum conditions in the research group of M. Rettenmayr (Otto Schott Institute of Materials Research, Friedrich-Schiller-University Jena, Germany). The ingot was re-melted several times and homogenized at $1000^\circ$C for $17$~hours. 
	
	\begin{figure}[h]
		\includegraphics[scale=.10]{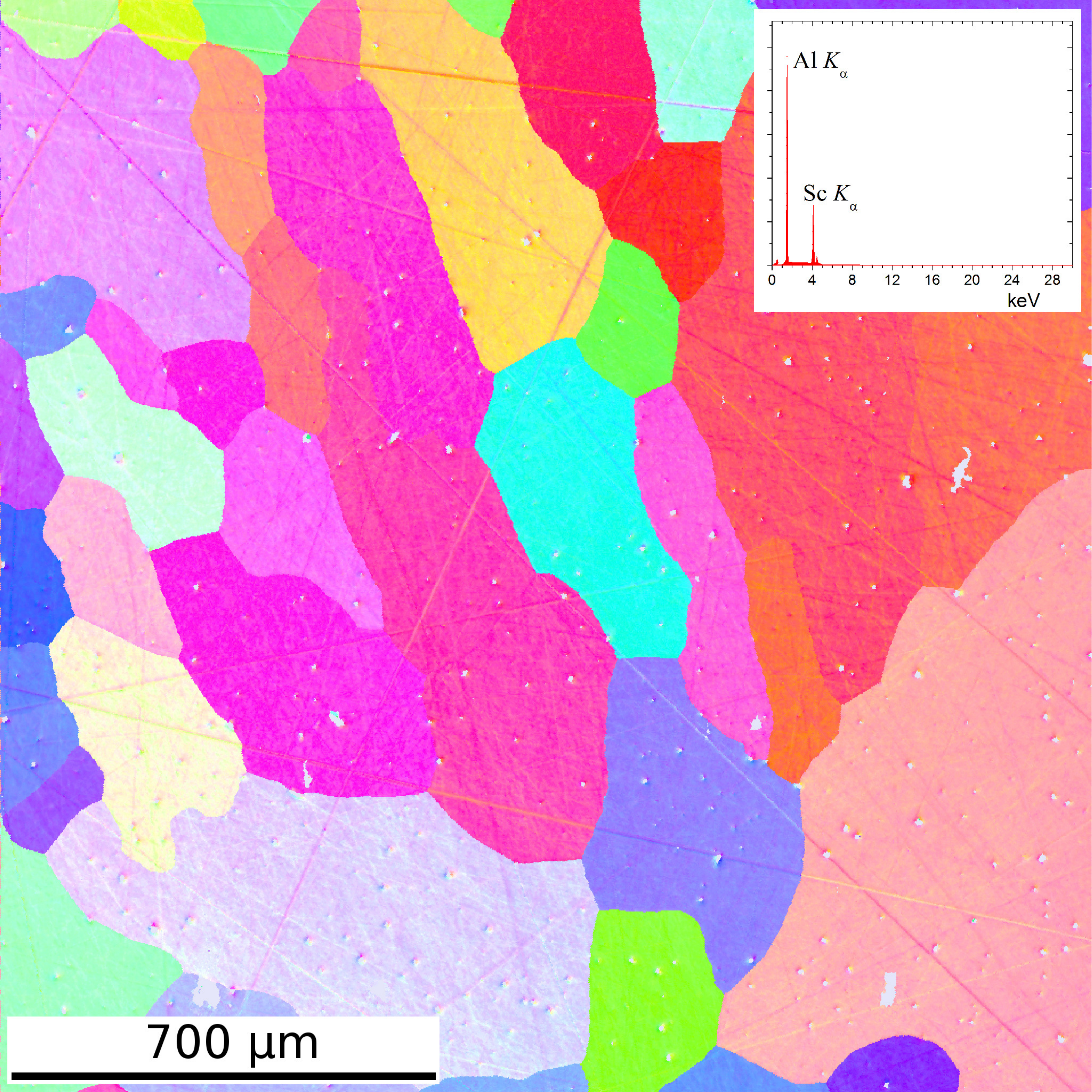}
		\caption{\label{EBSD}(Color online) Orientation imaging microscopy using EBSD analysis of the Al$_3$Sc sample. The recorded EDX spectrum is shown as inset and it reveals the presence of solely Al and Sc atoms.}
	\end{figure}
	
	Cube-shaped samples of size $3\times3\times3$ mm$^3$ were cut by spark erosion, etched, and polished. Low-temperature ($2$ to $400$~K) heat capacity measurements were performed by a Physical Property Measurement System (PPMS$^{\rm TM}$, Quantum Design Inc.). The microstructure of the cast and annealed Al$_3$Sc sample was checked using orientation imaging microscopy via an electron back-scatter diffraction (EBSD). A large area of $1500 \times 1500 \mu{\rm m}^2$ was scanned with a step of $2\mu {\rm m}$ (Fig.~\ref{EBSD}). The points with the confidence index smaller than 0.1 were excluded. The sample composition and homogeneity were controlled by an EDX analysis (inset Fig.~\ref{EBSD}). Small inclusions correspond to Al$_2$Sc particles and their volume fraction is estimated to be lower than 1\%.
	
	\section{Results and Discussion}
	
	\begin{figure*}
	\subfloat[Al\label{sfig:cp_kb_al}]{%
		\includegraphics[scale=.31]{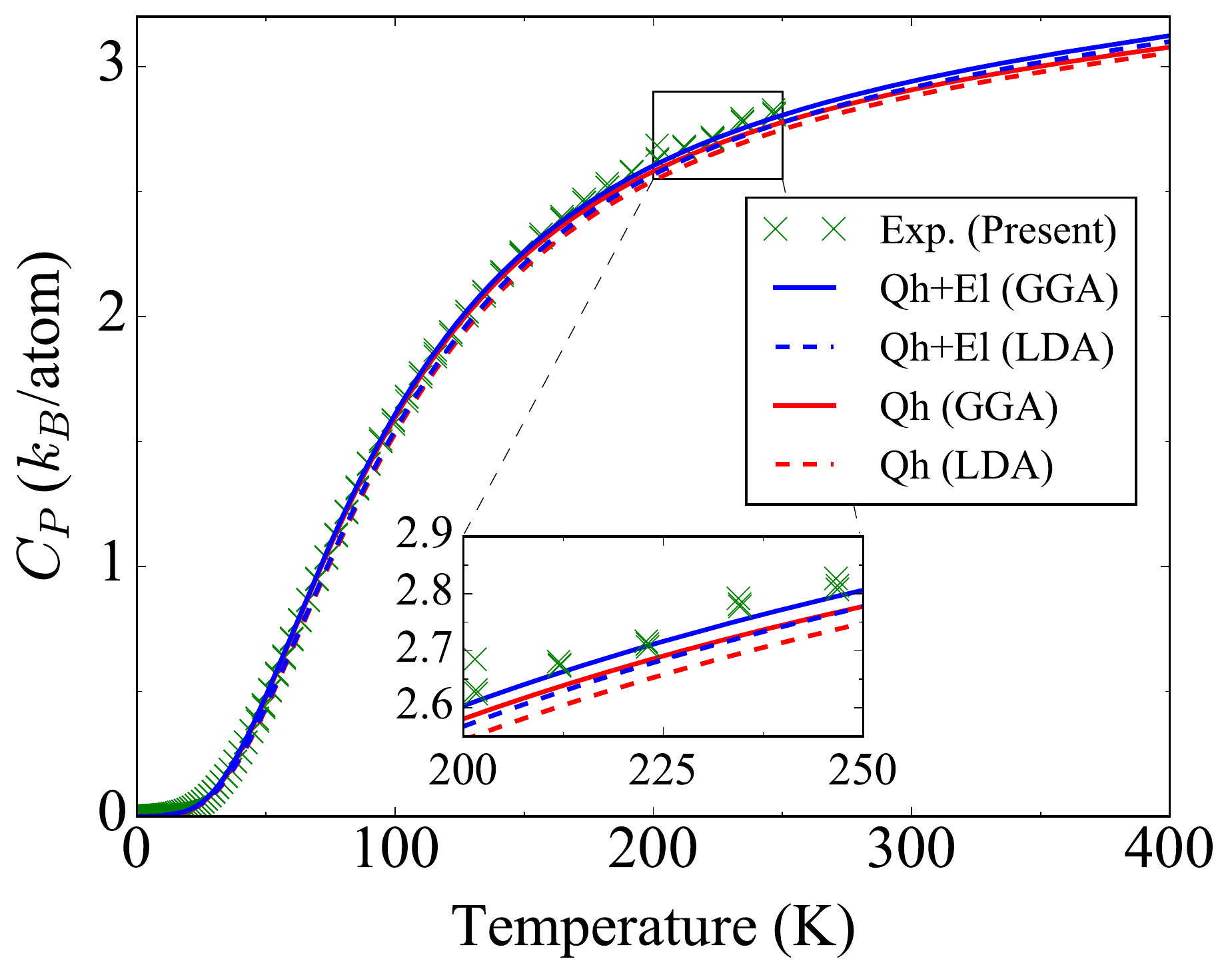}%
	}\hfill
	\subfloat[Al\label{sfig:cp_by_T3_al}]{%
		\includegraphics[scale=.305]{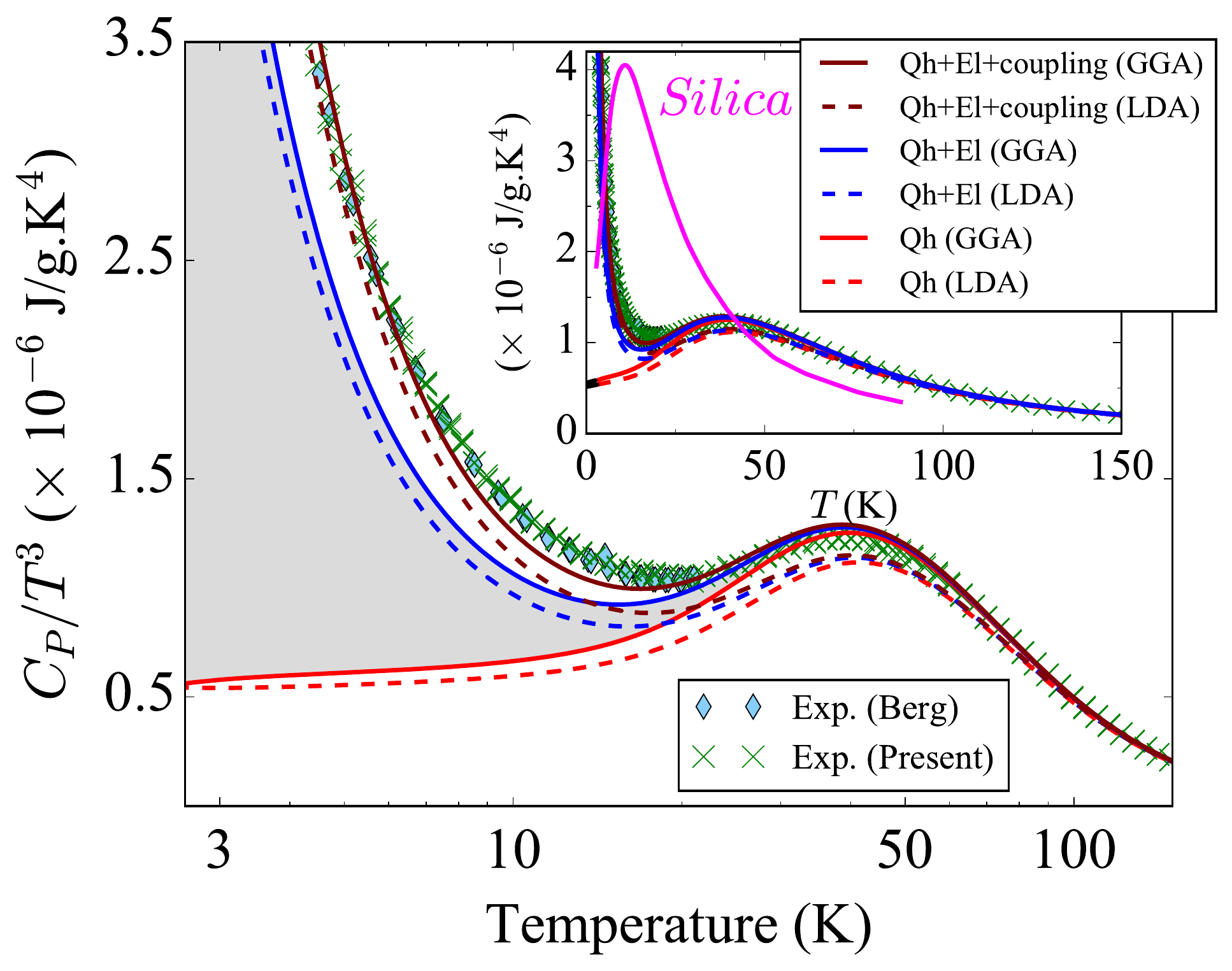}%
	}\hfill
	\subfloat[Al\label{sfig:sommerfeld_al}]{%
		\includegraphics[scale=.31]{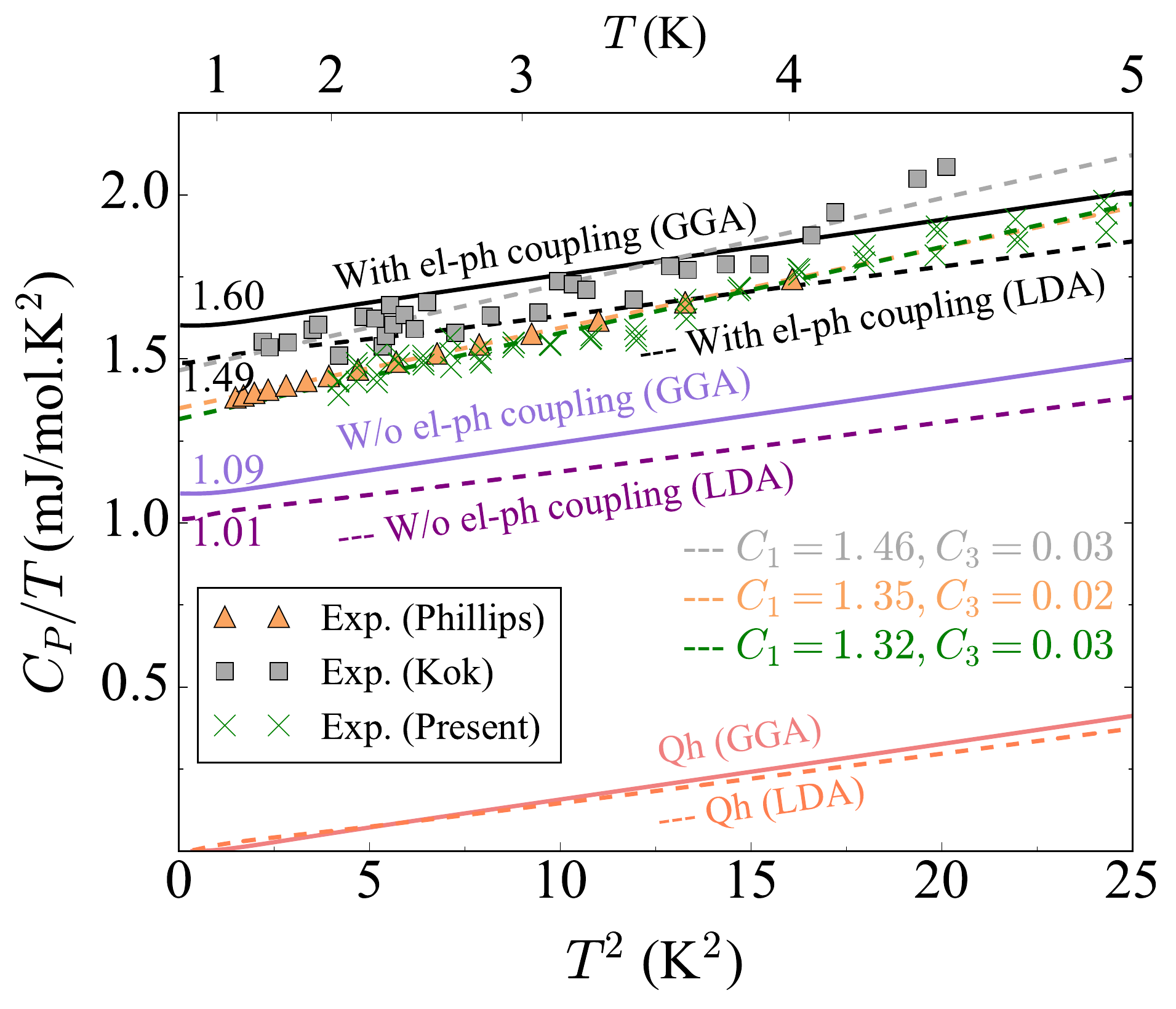}%
	}\vskip -0.9em
	\subfloat[Al$_3$Sc\label{sfig:cp_kb_al3sc}]{%
		\includegraphics[scale=.31]{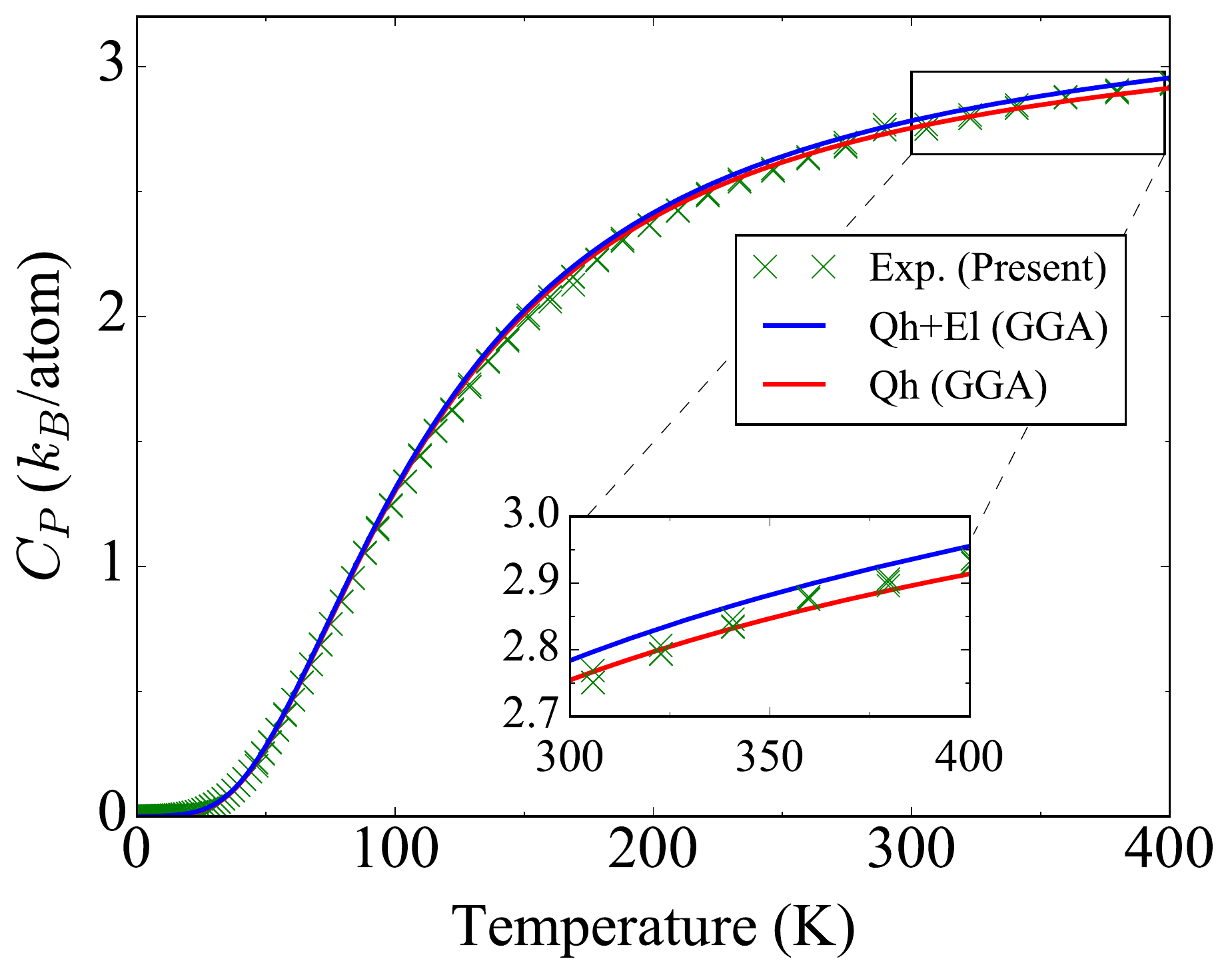}%
	}\hfill
	\subfloat[Al$_3$Sc\label{sfig:cp_by_T3_al3sc}]{%
		\includegraphics[scale=.31]{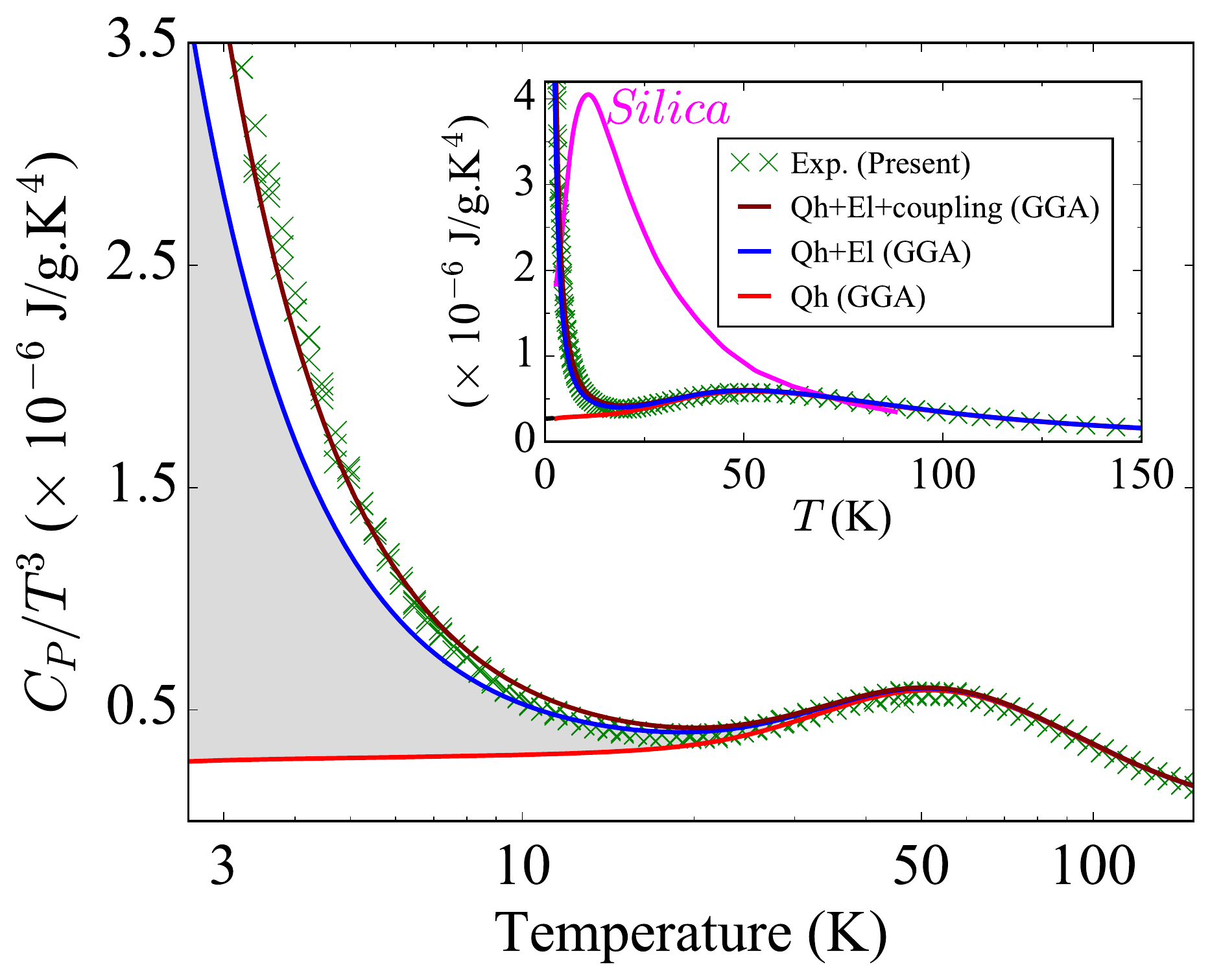}%
	}\hfill
	\subfloat[Al$_3$Sc\label{sfig:sommerfeld_al3sc}]{%
		\includegraphics[scale=.31]{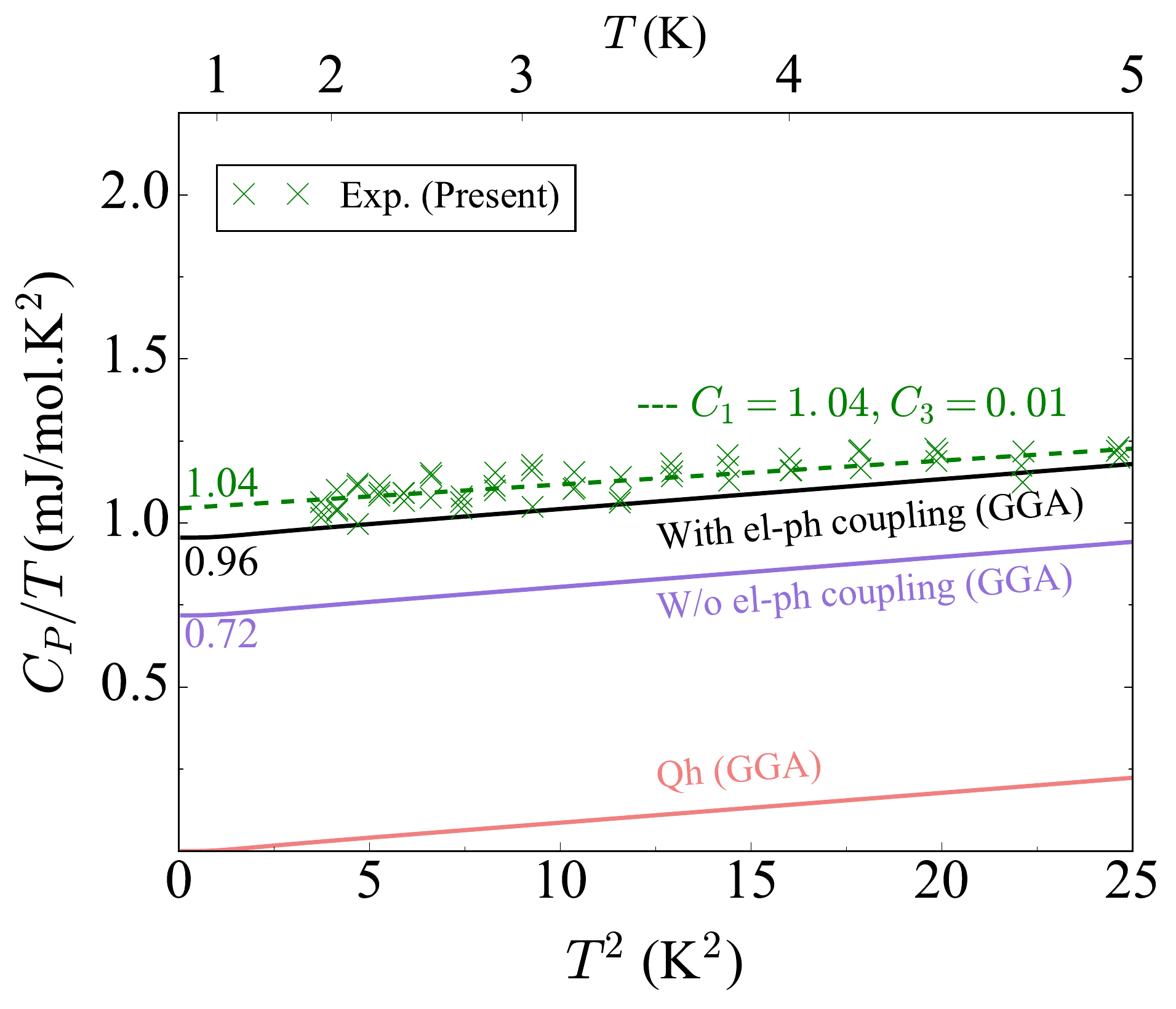}%
	}
	\caption{(Color online) Isobaric heat capacity ($C_P$) plots for Al and Al$_3$Sc. (a) and (d) Measured ($\color{OliveGreen} \times$ symbols) and calculated (solid lines: GGA, dashed lines: LDA) $C_P$ for Al and Al$_3$Sc in $k_B/\text{atom}$ respectively. The labels \textbf{Qh} and \textbf{El} correspond to the quasiharmonic and  electronic contributions, respectively. The inset provides a zoom in to highlight the electronic contribution. 
	(b) and (e) Renormalized $C_P/T^3$ curves for Al and Al$_3$Sc in J/g$\cdot$K$^4$. The filled $\color{CornflowerBlue} \blacklozenge$ symbols are experimental data for Al taken from Ref.~\onlinecite{CpAl_Berg}. The inset shows $C_P/T^3$ up to 150 K on a linear T axis. The magenta curve labeled \textit{Silica} is the renormalized $C_P/T^3$ for amorphous silica taken from Ref.~\onlinecite{Boson1} for comparison with the current work.
	(c) and (f) Plots of $C_P/T$ in mJ/mol$\cdot$K$^2$ versus $T^2$ to estimate the coefficients $C_1$ and $C_3$ for Al and Al$_3$Sc. The upper twin $x$-axis shows the absolute temperature in K. The solid (dashed) black and purple lines correspond to the GGA (LDA) calculations without (W/o) and with electron-phonon (el-ph) coupling respectively. The filled ($\color{orange} \blacktriangle$ and $\color{gray} \blacksquare$) symbols are the experimental data for Al taken from Refs.~\onlinecite{CpAl_Phillips,CpAl_Kok} respectively. The gray, green and orange dashed lines are the linear fits to the available low temperature experiments represented by symbols in the same color. The colored numbers written next to the $y$-axes are the $y$-intercepts ($C_1$) corresponding to different calculated curves. The coefficients $C_1$ and $C_3$ obtained from the fit to the available experiments are written in respective colors. }
	\label{cp_kb}
\end{figure*}
	
	In Figs.~\ref{sfig:cp_kb_al} and \ref{sfig:cp_kb_al3sc}, we present our measured (green crosses) and calculated (blue lines) $C_P$ data for Al and Al$_3$Sc.  We find a very good agreement over the whole temperature range. The good agreement with experiment is consistent with the small deviations between LDA and GGA results, since these deviations have been empirically found~\cite{Previous7} to provide for this quantity a confidence interval for the  predictive power (or accuracy) of the DFT calculations.

The DFT calculations allow by construction a  separation into the various free energy contributions, whereas the experimental analysis gives only access to the total response of the material. Specifically, the red curves in Figs.~\ref{sfig:cp_kb_al} and \ref{sfig:cp_kb_al3sc} correspond to the vibrational part as described within the quasiharmonic (\textbf{Qh}) approximation. The blue curves (\textbf{Qh}+\textbf{El}) show the impact of the electronic contribution which is small and of the same order as the tiny difference between theory and experiment (see inset in Fig.~\ref{sfig:cp_kb_al}). A similarly small electronic effect has been reported in previous DFT studies for  Al~\cite{Previous8,Previous12,Previous13} and Al alloys~\cite{CpAlAlloys}. Since the electronic contribution linearly decreases with temperature T, one would expect that it becomes even less significant at lower temperatures. 
	
	\subsection{Significance of the electronic contributions}
	
	\begin{figure}[thb]
		\vspace{-7mm}
		\includegraphics[scale=.42]{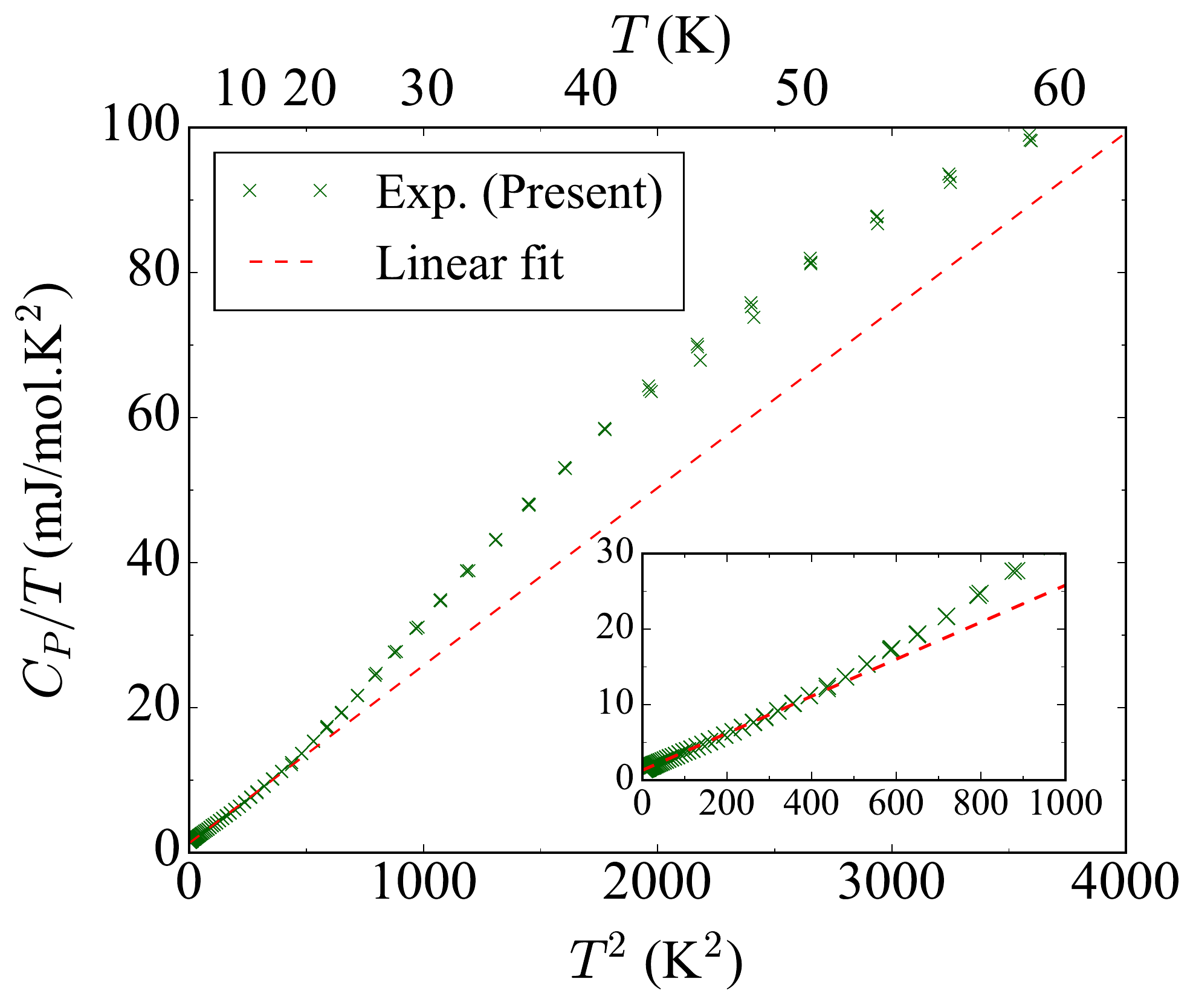}
		\caption{\label{linearity}(Color online) (Main) Linear fit (red dashed line) to the present experiments in the representation given by Eq.~(\ref{eq:cp_by_T}) up to approx.~60 K. The upper x-axis represents the absolute temperature $T$. (Inset) Zoom of the low temperature region up to  $T \approx 32$ K ($T^2 = 1000$ K).}
	\end{figure}
	
	Due to the $T^3$ decrease of the vibrational heat capacity towards low temperatures, the standard representation chosen in the plots in Figs.~\ref{sfig:cp_kb_al} and \ref{sfig:cp_kb_al3sc} is not the most suitable way to visualize and discuss features in the low temperature ($< 50$ K) regime. 
To analyze the validity of the Debye approximation, the heat capacity is often (e.g., in the literature for thermoelectric materials) renormalized with $T^{-3}$.  A perfect Debye dependence (Eq. \ref{eq:Debye}) would then result in a constant temperature independent behavior. 
Figures~\ref{sfig:cp_by_T3_al} and \ref{sfig:cp_by_T3_al3sc} show that the vibrational part (red curves) exhibits only a weak temperature dependence. The small deviations from $C_P \propto T^{3}$ are discussed in the next section. For reasons given in Sec.~\ref{sc:CompDetails}, the heat capacity is not plotted below 2 K. 
We note that the experimental measurements do not follow the constant behavior below $T\approx20$ K, but rather increase sharply, therewith indicating the importance of non-vibrational, i.e.~electronic excitations.

	For this reason, experimental measurements are commonly fitted to the sum of two contributions to describe the temperature dependence of the heat capacity in the low temperature regime~\cite{KittelBook,IIScBook}, 
	\begin{equation}
	\label{eq:CP1}
	C_P (T) = C_1 T + C_3 T^3.
	\end{equation} 
	The structure of this expression is motivated by the competition of electronic and vibrational contributions, as  captured by the Sommerfeld model (\ref{eq:Sommerfeld}) and the Debye model (\ref{eq:Debye}), respectively.   
	A straightforward approach to test the validity of Eq.~(\ref{eq:CP1}) is to perform another renormalization of the heat capacity data, namely by plotting  
	\begin{equation}
	\label{eq:cp_by_T}
	C_P/T = C_1  + C_3 T^2,
	\end{equation} 
	versus $T^2$. As shown in Fig.~\ref{linearity}, the experimental data show  in this representation  a constant slope up to approx.~20 K, i.e., up to this temperature the heat capacity is exclusively captured by a $T$ and $T^3$ dependence.
	
An advantage of the ab initio methodology is that it allows 
 a separate discussion of the different contributions to the heat capacity as shown in Figs.~\ref{sfig:sommerfeld_al} and \ref{sfig:sommerfeld_al3sc}. 
If only lattice vibrations contributed to the measured heat capacity at these temperatures (red lines), $C_P/T$ would pass through the origin. As already indicated in Sec.~\ref{sc:Vibrational} also the impact of a finite electronic temperature on the phonons will not change this behaviour. The same applies to anharmonic lattice vibrations, since our previous calculations~\cite{Previous8} for Al and our more recent results\cite{Upcoming_paper} for Al$_3$Sc only show a noticeable impact to the heat capacity close to the melting point. 
	
	The non-zero intercept with the $y$-axis in Figs.~\ref{sfig:sommerfeld_al} and \ref{sfig:sommerfeld_al3sc}, which determines $C_1$ in (\ref{eq:cp_by_T}), therefore, arises from  contributions linear in $T$. These are the electronic contribution and the electron-phonon coupling (\ref{eq:epcoupling}). We will first focus on the electronic contribution. 
The values of the coefficient $C_1$ obtained on the one hand by fitting the experimental data to Eq.~(\ref{eq:cp_by_T}) and on the other hand by computing the electronic contribution to the heat capacity are 1.04 and 0.72 mJ mol$^{-1}$K$^{-2}$ for Al$_3$Sc and 1.32-1.46 and 1.09 mJ mol$^{-1}$K$^{-2}$ for Al, respectively. The spread in the experimental data for the $C_1$ value of Al reflects the scatter in the available literature\cite{CpAl_Phillips,CpAl_Kok}. 
The sizeable difference between the experimental and theoretical $C_1$ indicates that  electron-phonon coupling in Eq.~(\ref{eq:epcoupling}) cannot be neglected. Thus, $C_1$ is not only determined by the Sommerfeld coefficient $\gamma$, but also by the coupling parameter $\lambda$.  
	For Al (Fig.~\ref{sfig:sommerfeld_al}), we have therefore computed the electron-phonon coupling and have determined the value 0.47 for $\lambda$, in good agreement with the reported $\lambda$ values for Al that range from 0.38 to 0.45\cite{SS96,Lambda_Al2,Lambda_Al3}. Using our $\lambda$ value in Eq.~(\ref{eq:epcoupling}) we get $C_1=1.6$ mJ mol$^{-1}$ K$^{-2}$ . This result agrees well with the experimental values (1.32-1.46 mJ mol$^{-1}$ K$^{-2}$)\cite{CpAl_Phillips,CpAl_Kok}. 
In the case of Al$_3$Sc (Fig.~\ref{sfig:sommerfeld_al3sc}) the calculation yields $\lambda=0.33$ (for Al$_3$Sc there exists no published data) corresponding to $C_1=0.96$ mJ mol$^{-1}$ K$^{-2}$ and showing again a good agreement with the experimental value of 1.04 mJ mol$^{-1}$K$^{-2}$.
	
	We can now revisit Figs.~\ref{sfig:cp_by_T3_al} and \ref{sfig:cp_by_T3_al3sc}. Taking in addition to the vibrational contribution (red lines) also the electronic contribution (blue curves) into consideration, we  observe that the \textit{ab initio} determined heat capacity agrees qualitatively as well as quantitatively well with the experiments. Based on the insights obtained with Figs.~\ref{sfig:sommerfeld_al} and \ref{sfig:sommerfeld_al3sc}, we can now conclude that the experimental data in the $C_P/T^3$ plot  
below 5 K corresponds to a $1/T^2$ behaviour, i.e. $C_P \sim T$. This is to a large extend captured by the electronic contribution to $C_P$. Taking in addition the electron-phonon coupling into account (brown lines), further improves the agreement with experiment. The latter contribution affects $C_P$ typically in the temperature range between 0.1-4 K\cite{Gri76,LIN}. 
Due to the observed linearity in the $C_P/T$ versus $T^2$ representation in Fig.~\ref{linearity}, however, we consider a temperature range up to 20 K for the electron-phonon coupling in Figs.~\ref{sfig:cp_by_T3_al} and \ref{sfig:cp_by_T3_al3sc}. 
	
	\begin{figure*} 
		\includegraphics[scale=0.6]{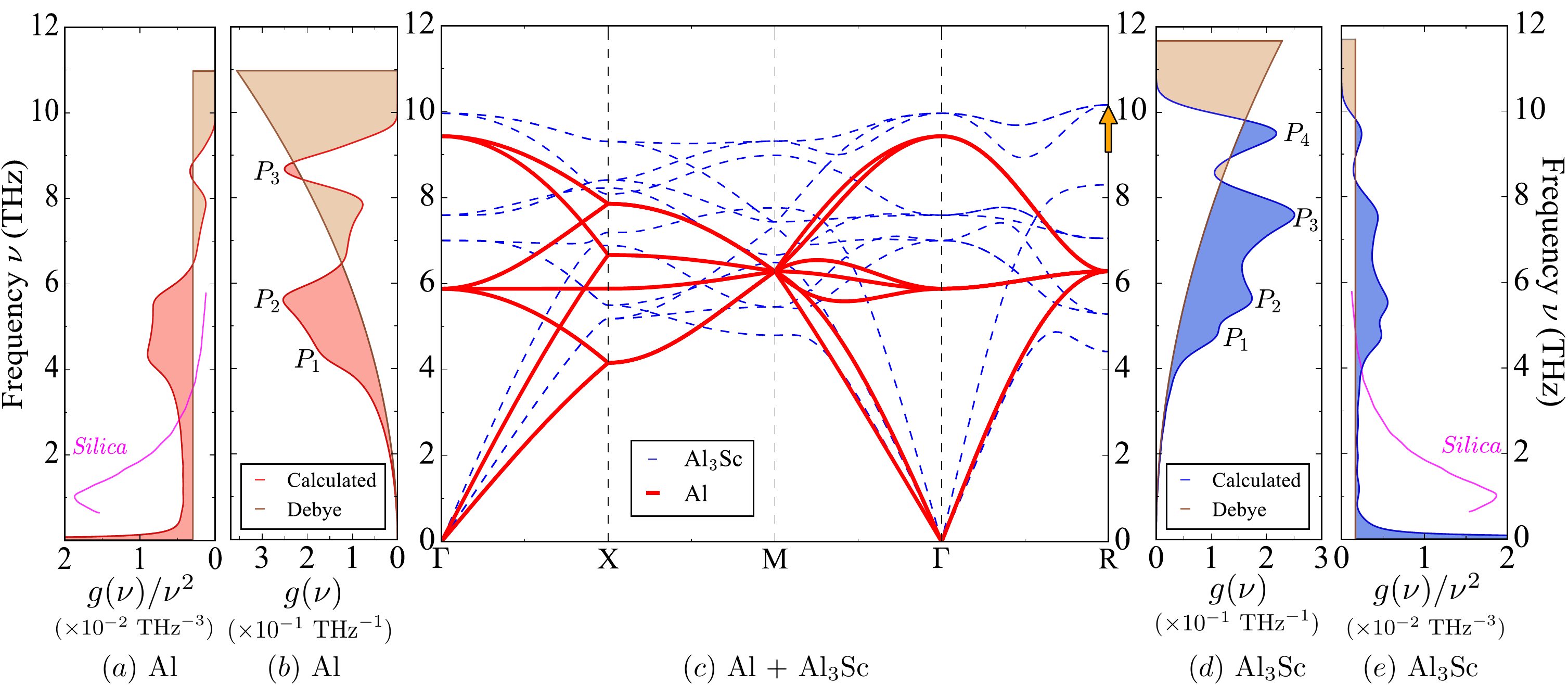}%
		\caption{(Color online) Calculated phonon dispersion curves of Al (solid red lines) and Al$_3$Sc (dashed blue lines) (c) plotted along a path in the BZ for Al$_3$Sc, (b and d) the corresponding vibrational density of states  $g(\nu)$ (VDOS) and (a and e) reduced VDOS  $[g(\nu)/\nu^2]$ for Al (left) and Al$_3$Sc (right). The brown curve represents the Debye VDOS ($g(\nu)\propto\nu^2$) up to the respective Debye frequencies. The magenta curve (taken from Ref.~\onlinecite{DOS_Silica}) represents the reduced VDOS for amorphous silica for comparison. The labels $P_1, P_2, P_3, P_4$ mark the observed peaks in the VDOS. The orange arrow at point R in (c) marks the selected phonon mode to study the effect of electronic temperature on the phonons (c.f. Sec.~\ref{sc:Vibrational}).}
		\label{dos_and_phonon}
	\end{figure*}
	
	As mentioned before, the fcc crystal structure is common to both Al and Al$_3$Sc phases where the Sc atoms replace the Al atoms on the corner sublattice sites resulting in the L1$_2$ Al$_3$Sc phase. Comparing the above discussed features for Al and Al$_3$Sc (Fig.~\ref{cp_kb}: all subplots are provided on the same scales), we observe that the impact of the electronic contribution and the electron-phonon coupling strength is reduced compared to that of pure Al. This is in contrast to  estimates in the literature, according to which the coupling constant $\lambda$ of an alloy can be approximated by the concentration-weighted average of the coupling constants of the individual elements\cite{WSB16,Sac80} (compare $\lambda=0.33$ for Al$_3$Sc with 0.47 for Al and 0.68 for Sc\cite{SA12}). 
The reason why this approach fails for Al$_3$Sc is the reduced density of states at the Fermi level in Al$_3$Sc compared to that of the pure metals and in particular Al. This reduces the number of partially occupied electronic states and thus the electronic entropy. The reduced density is the origin and driving force behind the energetic stability of this intermetallic compound, i.e., the failing of the averaging approach is a direct consequence of the fact that the electronic structure of the alloy cannot be regarded as small perturbation compared to the pure compounds.
 Since the same argument is expected to apply to any intermetallic compound, the average concept developed for alloys\cite{Sac80,WSB16} cannot be extended to ordered compounds.
 	
	\subsection{Non-Debye behaviour in the heat capacity}
	In addition to features below 20 K,  we observe in the renormalized heat capacity ($C_P/T^3$) plots (insets in Figs.~\ref{sfig:cp_by_T3_al} and \ref{sfig:cp_by_T3_al3sc}) maxima at approx.~40 K for Al and 50 K for Al$_3$Sc. At these temperatures the electronic contribution is already negligible and the Debye model would result in a constant line, i.e., the origin of these maxima must be related to an excess vibrational density of states (VDOS).
Such maxima in $C_P/T^3$ are extensively discussed in the amorphous materials community~\cite{Boson1,Boson2,Boson3,Boson4,Boson5} (e.g., for silica) and are called \textit{boson peak}. In contrast to the maxima we observe here, the peaks found for amorphous materials are much sharper (magenta curve with a maxima at around 15 K in insets of Figs.~\ref{sfig:cp_by_T3_al} and \ref{sfig:cp_by_T3_al3sc}) and are associated with localization phenomena (e.g., short range order) that shift the van Hove singularities to the low-frequency regime (close to 1 THz).
	
	In the crystalline and perfectly ordered materials investigated by us, the non-Debye behaviour in Figs.~\ref{sfig:cp_by_T3_al} and \ref{sfig:cp_by_T3_al3sc} has a different physical origin~\cite{BosonCrystalline1,BosonCrystalline2,BKA16}. To show this, we first provide the phonon spectra of Al and Al$_3$Sc (Fig.~\ref{dos_and_phonon}c),  plotted in the same BZ belonging to the L1$_2$ phase. 
The spectra are used to calculate the VDOS of Al and Al$_3$Sc (Figs.~\ref{dos_and_phonon}b and ~\ref{dos_and_phonon}d). 
In addition, we provide the Debye VDOS\cite{Maradudin} (brown solid curves in Figs.~\ref{dos_and_phonon}b and ~\ref{dos_and_phonon}d) as $g_D(\nu) = 3\nu^2/\nu_D^3$ with the Debye frequency $\nu_D = 2\pi(9\rho_{at}/4\pi)^{1/3}(2c_t^{-3}+c_l^{-3})^{-1/3}$. Here, $c_t$ and $c_l$ are the transverse and the longitudinal velocities of sound for metals calculated from the corresponding acoustic phonon branches close to the $\Gamma$ point and $\rho_{at}$ is the atomic concentration.  
Comparing the calculated and the Debye VDOS (in Figs.~\ref{dos_and_phonon}b and ~\ref{dos_and_phonon}d), we notice two profound peaks ($P_1$ and $P_2$) between 4-6 THz as well as additional peaks at higher frequencies that are characterized by an excess DOS over the Debye value. At the same time, Figs.~\ref{dos_and_phonon}a and ~\ref{dos_and_phonon}e compare the reduced VDOS ($g(\nu)/\nu^2$) of Al and Al$_3$Sc with that of silica for which we notice a sharp profound peak close to 1 THz.
	
\begin{figure}[tbh]
		\includegraphics[scale=.40]{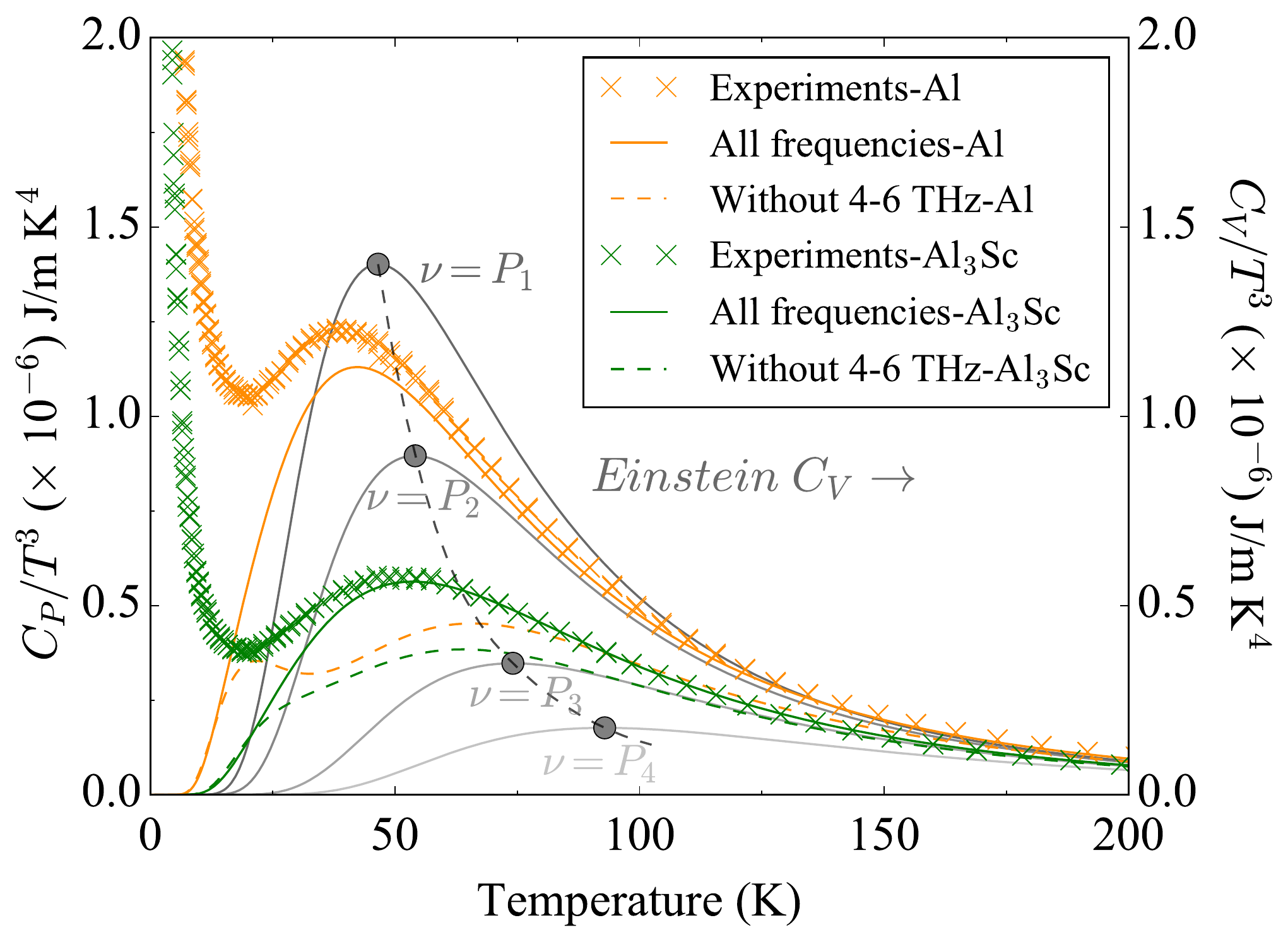}
		\caption{(Color online) Effect of excluding the 4-6 THz frequency band on the low temperature peak (orange and green dashed curves) in the renormalized heat capacity $C_V/T^3$ plot (right y-axis) for Al and Al$_3$Sc. The solid orange and green curves correspond to $C_V/T^3$ considering all the frequency modes in the phonon dispersion. The fading gray curves are the Einstein $C_V/T^3$ calculated at frequency $\nu$ corresponding to the peaks $P_1, P_2, P_3$ and $P_4$ in the VDOS of Al$_3$Sc (Fig.~\ref{dos_and_phonon}d) respectively. The dashed grey curve joins the maxima (grey circles) of the Einstein $C_V/T^3$ curves. Note that theory is given as $C_V/T^3$, whereas experimental data are given as $C_P/T^3$, resulting in an insignificant difference. }
		\label{fig5}
	\end{figure}

	It is difficult to experimentally resolve the effect the peaks have onto the shape of the $C_P/T^3$ curves, since the features are not very pronounced. In a theoretical investigation, we can analyze the heat capacity after extracting the frequency band of 4-6 THz. We use a representation of the BZ by 64 exact $q$ points (corresponding to lattice modulations fitting into a $4\times4\times4$ supercell) for this purpose, in order to achieve a well-defined weighting of the BZ. 
	As illustrated in Fig.~\ref{fig5} (dashed orange curve), the procedure suppresses the maximum in $C_P/T^3$ for Al at T $\approx$ 40 K and creates a new relatively flat maximum at a slightly higher temperature. Flat modes in the phonon dispersion curve at higher frequencies (e.g., $P_3$) are responsible for the remaining maximum. 
	
	
	To see how the addition of Sc affects the non-Debye behaviour, we compare the phonon dispersions of pure Al with Al$_3$Sc (Fig.~\ref{dos_and_phonon}c): The phonon branches in Al$_3$Sc are higher in frequency than in Al. This is attributed to the stiffer nature of the Al-Sc bonds as compared to Al-Al bonds due to a strong hybridization between Sc $d-$ and Al $p-$ electron states\cite{OzoAsta2001solvus} and also reflected by a higher bulk modulus value of 85.2 GPa for Al$_3$Sc in contrast to 74.6 GPa for Al. In addition, we observe a different relevance of flat optical branches in particular above 6 THz, yielding the peaks $P_3$ and $P_4$ in the VDOS of Al$_3$Sc (Fig.~\ref{dos_and_phonon}d). To check the impact of these peaks onto the maxima in the heat capacity of Al$_3$Sc, we repeated the procedure of neglecting the frequency band of 4-6 THz (which lead to $P_1$ and $P_2$). As shown in Fig.~\ref{fig5} (dashed green curve), we again observe a significant reduction in the maximum with a shift towards higher temperature as was also observed in the case of Al. Hence, the peak $P_3$ is less relevant than $P_1$ and $P_2$. 
	
	To evaluate the importance of these peaks in the VDOS, we use the Einstein model in which each phonon branch is effectively replaced by a constant single frequency $\nu_{\rm E}$ and the VDOS exhibits a single discrete peak. The Einstein heat capacity is given by~\cite{IIScBook},
	\begin{equation}
	C_V = 3 N k_B \frac{{\rm e}^{\frac{h\nu_{\rm E}}{k_BT}}}  {\left\{ {\rm e}^{\frac{h\nu_{\rm E}}{k_B T}}-1\right\}^2} \frac{(h\nu_{\rm E})^2}{(k_B T)^2}
	\end{equation}
	where $N,k_B, h$ are the number of atoms, the Boltzmann and the Planck constant, respectively. The critical temperature at which the heat capacity has a local maximum as per the Einstein model is $T_E (\nu) = h\nu_\text{E}/4.928 k_B$. 
	In Fig.~\ref{fig5} (fading gray curves), we plot the Einstein $C_V/T^3$ curves for the frequencies corresponding to $P_1, P_2, P_3$ and $P_4$ of Al$_3$Sc (Fig.~\ref{dos_and_phonon}d). With increasing frequency ($\nu_{P_1}<\nu_{P_2}<\nu_{P_3}<\nu_{P_4}$), the maximum is observed to flatten and to shift towards higher temperatures. This trend agrees with our foregoing comparison of maxima for Al and Al$_3$Sc: The peaks in the Al$_3$Sc VDOS are at higher frequencies (owing to stiffer phonons) than those for Al. Consequently, the maximum is flatter and observed at a slightly higher temperature than Al.
	Therefore, the non-Debye behaviour of crystals is mainly determined by the low-lying flat regions in the phonon spectrum, i.e. the region of 4-6 THz in Al and Al$_3$Sc is responsible for the maxima around 40-50 K in the heat capacity. Comparing the heat capacities of Al and Al$_3$Sc (Figs.~\ref{sfig:cp_by_T3_al} and \ref{sfig:cp_by_T3_al3sc}), one can notice that the maximum is less pronounced for Al$_3$Sc and occurs at higher temperature.

	\section{Conclusions}
	
	Based on the insights gained in this study, we demonstrate: (i) The predictive capability of finite-temperature DFT in accurately describing the competition between electronic and vibrational heat capacity contributions in Al and Al$_3$Sc at temperatures well below the Debye temperature, if highly converged calculations are performed. (ii) The importance of contributions beyond the adiabatic approximation, in particular of electron-phonon coupling for the correct determination of the electronic heat capacity at low temperatures. (iii) The connection of flat branches in the phonon spectrum and the observed maxima in the low-temperature heat capacity measurements that are not captured by the Debye model.
	
	The availability of highly accurate DFT computed heat capacities including the partition in the various entropic contributions allowed a detailed analysis about the validity of approximate but commonly applied physical models such as Sommerfeld theory or Debye model. To visualize and discuss the relevant low temperature features we considered various renormalized transformations. 
	
	The comparison of Al and Al$_3$Sc reveals that the observed deviations from the Debye behaviour at low temperatures are particularly strong in the pure element Al. The addition of 25 at.\% Sc to Al reduces the fcc symmetry due to the L1$_2$ superstructure. Despite the higher structural complexity, however, the low-temperature features in the heat capacity become less pronounced. This is related to the higher stiffness of the Al-Sc bonds and the resulting increase of the Debye temperature. In an Al-Sc alloy that phase separates into a solid solution and Al$_3$Sc precipitates, the thermodynamic properties at low temperatures will therefore be determined by the Al-rich matrix.   While this study focussed on two example systems -- fcc Al and Al$_3$Sc -- all discussions and considerations regarding the methodological approaches and the analysis of the data are general and can be directly transferred to other materials and compounds.   
	
	\begin{acknowledgments}
		The authors thank Dr.~S.~Lippmann and Prof.~M.~Rettenmayr (Otto Schott Institute of Materials Research, Friedrich-Schiller-University Jena, Germany) for fruitful discussions and support in preparation of Al$_3$Sc samples. The authors are also grateful to Prof.~M.~W.~Finnis and Dr.~X.~Zhang for valuable discussions and suggestions. Financial support from the Deutsche Forschungsgemeinschaft (DFG) within the priority program SPP-1713 ``chemomechanics'' (research projects HI 1300/8-1 and DI 1419/7-1) is gratefully acknowledged.
	\end{acknowledgments}
	
	\newpage
	

\end{document}